\documentclass[prl, twocolumn, superscriptaddress]{revtex4}
\usepackage{graphicx,color}
\usepackage{amssymb}
\usepackage{amsmath}
\usepackage{bm}
\usepackage[colorlinks=true,linkcolor=blue,urlcolor=blue,citecolor=green]{hyperref}
\usepackage{tabularx}
\usepackage{multirow}
\usepackage{braket}
\usepackage{xcolor}
\usepackage{gensymb}
\usepackage{amsmath,amssymb}

\begin{document}

\title{Second Harmonic Hall Response in Insulators: Inter-band Quantum Geometry and Breakdown of Kleinman's Conjecture}
\author{Wen-Yu He}\thanks{hewy@shanghaitech.edu.cn}
\affiliation{State Key Laboratory of Quantum Functional Materials, School of Physical Science and Technology, ShanghaiTech University, Shanghai 201210, China}
\author{K. T. Law}\thanks{phlaw@ust.hk}
\affiliation{Department of Physics, Hong Kong University of Science and Technology, Clear Water Bay, Hong Kong, China}

\date{\today}

\begin{abstract}
The nonlinear Hall effect has recently garnered significant attention as a powerful probe of Fermi surface quantum geometry in metals. While current studies mainly focus on the nonlinear Hall response driven by quasi-static electric fields of low frequencies, the extension of the response to higher frequencies is another promising frontier, which introduces quantum geometry into inter-band transitions. Here, we demonstrate that a specific nonlinear Hall response, namely the second harmonic Hall (SHH) response, can arise from inter-band transitions. We establish the quantum geometric origin of the SHH response and show that inter-band quantum geometry dominates the SHH response when driven near inter-band resonance. Crucially, we find that the inter-band SHH response in insulators exhibits strong frequecy dispersion, manifesting the breakdown of Kleinman's conjecture in nonlinear optics. This connects the SHH response to the breakdown of Kleinman's conjecture and reveals that frequency dispersive insulators generally allow the SHH response. Furthermore, we predict a giant SHH susceptibility in gated strained bilayer graphene and propose that one can apply the polarization resolved second harmonic microscopy to detect the SHH response there.
\end{abstract}

\maketitle

\emph{Introduction}.--- The Hall effect in materials is a fundamental phenomenon that not only has extensive practical applications~\cite{EdinHall, Ramsden, Nagaosa, Cuizu} but also fosters a deeper understanding of topology, Berry curvature, and quantum geometry in electronic transport~\cite{Cuizu, Nagaosa, Klitzing, Thouless, QianNiu}. Recently, the Hall effect has been extended beyond the linear regime, unveiling the nonlinear Hall effect~\cite{LiangFu1, QianNiu02, Dixiao01, Shengyuan, Sodemann01, Sodemann02, Shengyuan02, Law01, Resta, Haizhou, Haizhou2, Noejung, Binghai1, YuanFang, Amit, Binghai2, Haizhou3, Ortix01, Narayan} as a powerful probe of quantum geometry in a series of metals~\cite{QiongMa, KinFai, Mandar, AGao, Weibo, PanHe01, HyunsooYang, TseMing, Yugui, NingWang2, NingWang, Zhiqiang, ZhiminLiao, Changgan, Loh, Weibo2, Berthold}. In these metals, a quasi-static electric field (in the low frequency range from 0 to a few Hz) couples to intra-band quantum geometric quantities such as Berry curvature dipole (BCD)~\cite{LiangFu1}, quantum metric dipoles~\cite{QianNiu02, Dixiao01, Shengyuan, Binghai2} and their higher moments~\cite{Law01,YuanFang}, generating a nonlinear voltage response transverse to the applied electric field. Given the driving electric field in the low frequency range, it has been established that the nonlinear Hall effect is governed by the quantum geometry on the Fermi surfaces~\cite{LiangFu1, QianNiu02, Dixiao01, Shengyuan, Binghai2, Sodemann01, Sodemann02, Shengyuan02,Law01,YuanFang, QiongMa, KinFai, AGao, Weibo}.

While the scenario of Fermi surface quantum geometry works well for the nonlinear Hall effect driven by quasi-static electric fields, a key question arises: how does the nonlinear Hall response evolve when the driving frequency increases to involve inter-band effects? In the inter-band processes, electrons from occupied bands get excited or virtually excited to unoccupied bands, activating frequency dependent inter-band quantum geometric contributions to the nonlinear response~\cite{Nagaosa02, Parker, Nagaosa03, Raquel01,Raquel02}. Notably, it is known that the second harmonic generation (SHG) can arise from electronic inter-band transitions irrespective of Fermi surface presence~\cite{Sodemann02, Noejung, Sipe2000, Tobias023, Amit02}. Crucially, the SHG transverse to the applied electric field constitutes a defining feature of the nonlinear Hall response~\cite{LiangFu1, QiongMa, KinFai, Mandar, AGao, Weibo, PanHe01, HyunsooYang, TseMing, Yugui, NingWang2, NingWang, Zhiqiang, ZhiminLiao, Changgan, Loh}. In Ref. \citenum{Noejung}, it has been found that inter-band BCD has the potential to induce a transverse SHG. However, the full explicit connection between the transverse SHG and inter-band quantum geometry has yet been established, leaving a critical gap in understanding the nonlinear Hall response that involves inter-band processes.

\begin{figure*}[t]
\centering
\includegraphics[width=0.9\textwidth]{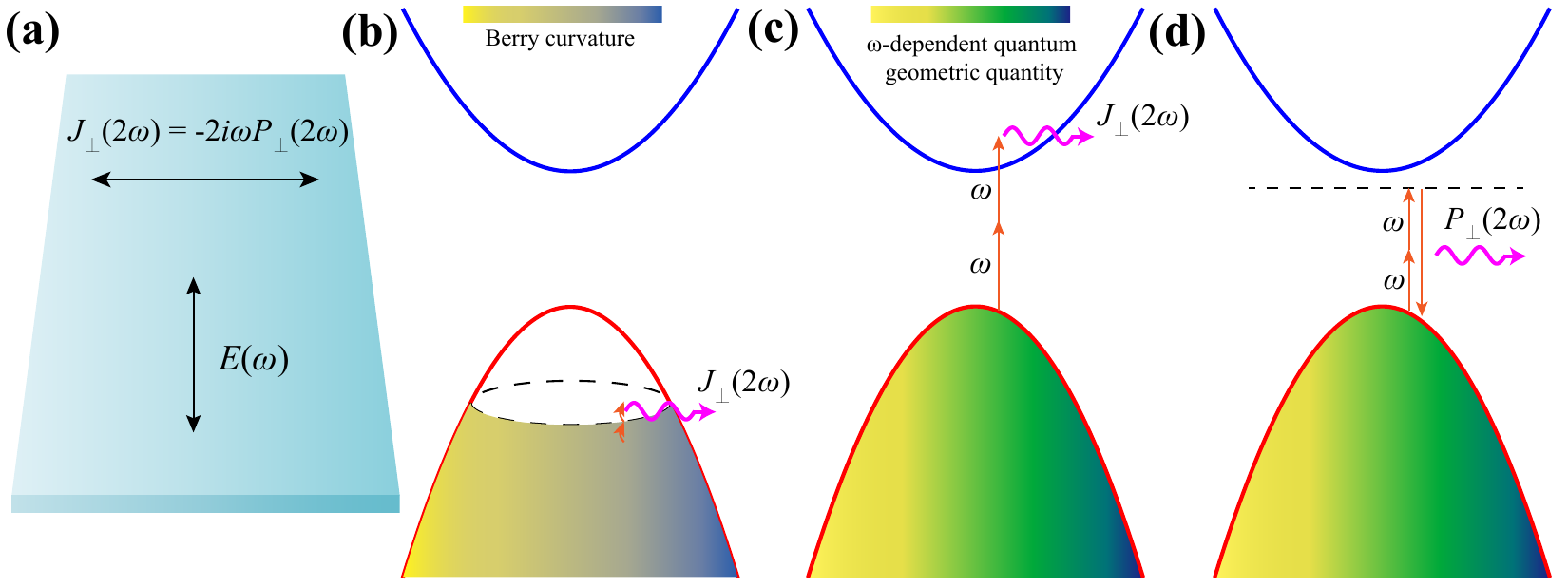}
\caption{Illustration of the SHH response involving inter-band processes. (a) The SHG transverse to the applied electric field serves as a distinctive signature of SHH response. (b) The BCD induced transverse SHG in a metal. The redistribution of electrons near the Fermi surface imbalances the Berry curvature and induces the SHH current response. In (c) and (d), the valence bands get fully occupied and the system is insulating. In the resonant regime in (c), electrons in the occupied bands are pumped onto the empty conduction bands, while in the slightly off-resonant regime in (d), electrons go through a virtual inter-band transition. In the inter-band processes, frequency dependent quantum geometric quantities in the occupied bands can give rise to a finite $\bm{J}_\perp\left(2\omega\right)$ in (c) and a finite $\bm{P}_\perp\left(2\omega\right)$ in (d) respectively. Both $\bm{J}_\perp\left(2\omega\right)$ and $\bm{P}_\perp\left(2\omega\right)$ manifest the SHH response.}
\label{fig1}
\end{figure*}

In this work, we resolve this problem by unraveling the quantum geometric origin of the transverse SHG, which we refer as the second harmonic Hall (SHH) response. As illustrated in Fig. \ref{fig1}, we demonstrate that 1) frequency dependent inter-band quantum geometry contributes significantly to the SHH response when inter-band transitions are enabled; 2) insultors can exhibit a pronounced SHH response when driven at frequency near inter-band resonance. Remarkably, the SHH response in insulators exhibits a significant frequency dispersion: it vanishes far off-resonance but get enhanced sharply near inter-band resonance. This behavior directly manifests the breakdown of Kleinman's conjecture in nonlinear optics~\cite{Boyd, Cotter, Kleinman}, where the overall permutation symmetry of the susceptibility indices breaks down due to the frequency dispersive properties of the material.

The physical mechanism of the SHH response involving inter-band transitions can be inferred from the schematic diagrams in Fig. \ref{fig1}. In a time reversal invariant metal (Fig. \ref{fig1} (b)), a quasi-static electric field redistributes electronic states near the Fermi surface, creating a net BCD that generates the SHH current~\cite{LiangFu1}. When the driving field induces inter-band transitions (Fig. \ref{fig1} (c)), electrons are excited from valence bands to conduction bands, enabling inter-band redistributions of electronic states. During the inter-band processes, frequency dependent inter-band quantum geometry modifies electrons' motion and induces an inter-band SHH current. Crucially, in the slightly off-resonance regime (Fig. \ref{fig1} (d)), virtual inter-band transitions~\cite{Raquel01} cause electronic states to vibrate in the direction transverse to the driving field, generating a SHH polarization $\bm{P}_\perp\left(2\omega\right)$. The inter-band SHH response illustrated in Fig. \ref{fig1} (c) and (d) can occur in both metals and insulators, but we focus here on insulators, which inherently exclude Fermi surface effects while retaining pure inter-band quantum geometry.

In the following, we first derive the explicit SHH conductivity via the density matrix formalism~\cite{Passos01, Sipe01, Sipe02, Ventura} and demonstrate that insulators can exhibit finite, frequency dispersive SHH response when driven near inter-band resonance. This is consistent with the breakdown of Kleinman's conjecture~\cite{Boyd, Cotter, Kleinman}, indicating that frequency-dispersive insulators are prime candidates for strong SHH response. We validate our theory by predicting a giant SHH susceptibility in bias-gapped Bernal bilayer graphene under uniaixal strain, with responses exceeding those of transition metal dichaocogenides by two orders of magnitude~\cite{Xiaodong, Soavi}. Finally, we propose that one can utilize the polarization resolved second harmonic microscopy~\cite{HuiZhao, Paula, YLi}to detect the SHH response in insulators.

\begin{table*}[t]
	\caption{The explicit expressions for the SHH conductivity. The SHH conductivity takes the form $\sigma^\perp_{qij}\left(2\omega;\omega,\omega\right)=\epsilon_{ljq}\sum_{n=1}^4\gamma^{\left(n\right)}_{il}\left(\omega\right)+\left(i\leftrightarrow j\right)$ with all the 4 terms of $\gamma^{\left(n\right)}_{il}\left(\omega\right)$ listed in the Table. The terms $\gamma^{\left(n\right)}_{il}\left(\omega\right)$ consist of integrals over the Brillouin zone. $\gamma^{\left(1\right)}_{il}\left(\omega\right)$ accounts for the Fermi surface contribution and the other terms $\gamma^{\left(n\right)}_{il}\left(\omega\right)$ with $n=2,3,4$ denotes the contribution from the inter-band transitions. In the integration, $\bm{A}_{ab,\bm{k}}=i\braket{u_{a,\bm{k}}|\partial_{\bm{k}}u_{b,\bm{k}}}$ is the non-Abelian Berry connection, $\hat{\partial}_{\bm{k}}$ denotes the generalized derivative operator, and $g^{\left(\nu\right)}_{ac,\bm{k}}$ with $\nu=1,2,3$, $g^{\left(4\right)}_{acc_1,\bm{k}}\left(\omega\right)$,  $g^{\left(5\right)}_{cc_1a,\bm{k}}\left(\omega\right)$ are the frequency dependent form factors. The subscripts $i,l=x,y,z$ are the indices of spatial directions. All the integrands of $\gamma^{\left(n\right)}_{il}\left(\omega\right)$ respect the gauge invariance, so they are referred as the frequency dependent quantum geometric quantities. More details about the generalized derivative operator and the frequency dependent form factors can be found in the Supplemental Material~\cite{Supp}.}
	
	\centering  
	{
		\begin{tabular}{c c }
			\hline \hline
			\rule{0pt}{3ex}
			 Contributing terms  & Frequency dependent quantum geometric quantities \\ [1ex]
			\hline
			\rule{0pt}{3ex}
			$\gamma^{\left(1\right)}_{il}\left(\omega\right)$ & $ \int_{\bm{k}}\sum_a\partial_{k_i}f_{a,\bm{k}}\sum_{c\neq a}\left(i\bm{A}_{ac,\bm{k}}\times\bm{A}_{ca,\bm{k}}\right)_lg^{\left(1\right)}_{ac,\bm{k}}\left(\omega\right)$ \\ [2ex] \rule{0pt}{2ex}
			$\gamma^{\left(2\right)}_{il}\left(\omega\right)$ & $ \int_{\bm{k}}\sum_{a,c\neq a}f_{a,\bm{k}}i\left[A^i_{ac,\bm{k}}\left(\hat{\partial}_{\bm{k}}\times\bm{A}_{ca,\bm{k}}\right)_l-\left(\hat{\partial}_{\bm{k}}\times\bm{A}_{ac,\bm{k}}\right)_lA^i_{ca,\bm{k}}+\partial_{k_i}\left(\bm{A}_{ac,\bm{k}}\times\bm{A}_{ca,\bm{k}}\right)_l\right]g^{\left(2\right)}_{ac,\bm{k}}\left(\omega\right)$  \\ [2ex]\rule{0pt}{2ex}
			$\gamma^{\left(3\right)}_{il}\left(\omega\right)$ & $\int_{\bm{k}}\sum_{a,c\neq a}f_{a,\bm{k}}\left[\left(i\bm{A}_{ac,\bm{k}}\times\bm{A}_{ca,\bm{k}}\right)_l\Delta E_{ac,\bm{k}}\partial_{k_i}\Delta E_{ac,\bm{k}}\right]g^{\left(3\right)}_{ac,\bm{k}}\left(\omega\right)$ \\[2ex]\rule{0pt}{2ex}
			 $\gamma^{\left(4\right)}_{il}\left(\omega\right)$ & $\int_{\bm{k}}\sum_{a,c\neq a,c_1\neq c,c_1\neq a}\left[f_{a,\bm{k}}g^{\left(4\right)}_{acc_1,\bm{k}}\left(\omega\right)-f_{c,\bm{k}}g^{\left(5\right)}_{cc_1a,\bm{k}}\left(\omega\right)\right]\left[A^i_{cc_1,\bm{k}}\left(i\bm{A}_{ac,\bm{k}}\times\bm{A}_{c_1a,\bm{k}}\right)_l-A^i_{c_1c,\bm{k}}\left(i\bm{A}_{ac_1,\bm{k}}\times\bm{A}_{ca,\bm{k}}\right)_l\right]$  \\ [2ex] 
						\hline 
	\end{tabular}}
	\label{table1}
\end{table*}

\emph{General theory for the SHH Response}.--- To study the SHH response, one needs to first extract the transverse component from the standard SHG conductivity. From the density matrix formalism~\cite{Passos01, Sipe01, Sipe02, Ventura}, we know that the SHG conductivity in a two or three dimensional system ($d=2, 3$) is derived to be~\cite{Supp}
\begin{widetext}
\begin{align}\label{2nd_harmonic_conduc}
\sigma_{qij}\left(2\omega;\omega,\omega\right)=&\frac{e^3}{2\hbar}\int_{\bm{k}}\sum_af_{a,\bm{k}}\left\{\left[\hat{D}_{i},\frac{1}{\hbar\bar{\omega}+\Delta E}\circ\left[\hat{D}_{j},\frac{1}{2\hbar\bar{\omega}+\Delta E}\circ\left[\hat{D}_q,\hat{H}_0\right]\right]\right]_{aa,\bm{k}}+\left(i\leftrightarrow j\right)\right\},
\end{align}
\end{widetext}
with $\hat{\bm{D}}$ being the covariant derivative operator~\cite{Passos01, Ventura,Supp}, $\hat{H}_0$ being the Hamiltonian, $\int_{\bm{k}}\equiv\int d\bm{k}/\left(2\pi\right)^d$, $f_{a,\bm{k}}$ denoting the occupation of the Bloch state $\ket{\psi_{a,\bm{k}}}$. Here $\left[\dots\right]$ means the commutator and $\left(\hat{O}\circ\hat{B}\right)_{ab}=\hat{O}_{ab}\hat{B}_{ab}$ is the Hadamard product. We have assumed that the applied electric field is adiabatically switched on~\cite{Passos01} so a phenomenological relaxation time is associated with the frequency: $\bar{\omega}=\omega+i\tau^{-1}$. In our calculations, the relaxation time is set to be $\hbar\tau^{-1}=1$ meV.

The SHG conductivity $\sigma_{qij}\left(2\omega;\omega,\omega\right)$ in Eq. \ref{2nd_harmonic_conduc} respects the intrinsic permutation symmetry $\sigma_{qij}\left(2\omega;\omega,\omega\right)=\sigma_{qji}\left(2\omega;\omega,\omega\right)$ that the two indices $i$ and $j$ are commutative~\cite{Boyd, Cotter}. In contrast, the output index $q$ is generally non-commutative with $i$ and $j$. From the transverse nature of the Hall response, we know that the SHH conductivity corresponds to the anti-symmetric part of $\sigma_{qij}\left(2\omega;\omega,\omega\right)$ when exchanging $q$ with $i$ and $j$. Thus, the SHH conductivity is constructed as
\begin{align}\nonumber\label{sigma_H01}
\sigma^\perp_{qij}\left(2\omega;\omega,\omega\right)=&\frac{1}{3}\left[2\sigma_{qij}\left(2\omega;\omega,\omega\right)\right.\\
&\left.-\sigma_{iqj}\left(2\omega;\omega,\omega\right)-\sigma_{jiq}\left(2\omega;\omega,\omega\right)\right].
\end{align}
It can be checked that $\sigma^\perp_{qij}\left(2\omega;\omega,\omega\right)$ satisfies the anti-symmetric requirement of the Hall conductivity while preserving the intrinsic permutation symmetry of $i$ and $j$~\cite{Supp}. This way of constructing the SHH conductivity is in the same spirit as the prescription introduced in Ref.~\citenum{Tsirkin}, and one can further extend it to higher orders. Now, by combining Eq. \ref{2nd_harmonic_conduc} and Eq. \ref{sigma_H01}, we are ready to present the complete expression for the SHH conductivity.

\emph{The SHH conductivity in insulators}.--- After explicit calculations performed in Supplemental Materials~\cite{Supp}, we obtain the SHH conductivity with the form
\begin{align}
\sigma^{\perp}_{qij}\left(2\omega;\omega,\omega\right)=&\epsilon_{ljq}\gamma_{il}\left(\omega\right)+\left(i\leftrightarrow j\right).
\end{align}
Here the Levi-Civita symbol $\epsilon_{ljq}$ is introduced to directly show the anti-symmetric property of $\sigma^{\perp}_{qij}\left(2\omega;\omega,\omega\right)$. The essential elements of $\sigma^{\perp}_{qij}\left(2\omega;\omega,\omega\right)$ are contained in the rank-2 pseudotensor $\gamma_{il}\left(\omega\right)=\sum_{n=1}^4\gamma_{il}^{\left(n\right)}\left(\omega\right)$, where the explicit expressions of $\gamma_{il}^{\left(n\right)}\left(\omega\right)$ are listed in Table \ref{table1}. It is clear that the integrand in $\gamma^{\left(n\right)}_{il}\left(\omega\right)$ is composed of a rank-2 pseudotensor multiplied by a frequency dependent form factor, where the rank-2 pseudotensor has the general structure of $v_i\left(\bm{v}_1\times\bm{v}_2\right)_l$ with the vectors $\bm{v}$, $\bm{v}_1$ and $\bm{v}_2$ selected from the non-Abelian Berry connection $\bm{A}_{ab,\bm{k}}$ and the $\bm{k}$-space gradient $\partial_{\bm{k}}$. Here we have assumed that the system respects the time reversal symmetry, but one can generalize the combinations of $\partial_{\bm{k}}$, $\bm{A}_{ab,\bm{k}}$ and $f_{a,\bm{k}}$ in Table \ref{table1} to the case without time reversal symmetry and higher order harmonic response. For the first term $\gamma^{\left(1\right)}_{il}\left(\omega\right)$, the gradient $\partial_{\bm{k}}$ acts on the distribution function $f_{a,\bm{k}}$, so $\gamma^{\left(1\right)}_{il}\left(\omega\right)$ purely accounts for the contribution from Fermi surfaces. In the low frequency limit $\omega\rightarrow0$, $\gamma^{\left(1\right)}_{il}\left(\omega\right)$ reduces to the BCD~\cite{LiangFu1}. The other terms $\gamma^{\left(n\right)}_{il}\left(\omega\right)$ with $n=2,3,4$, however, are more general combinations of $\partial_{\bm{k}}$, $\bm{A}_{ab,\bm{k}}$ and $f_{a,\bm{k}}$ that exclude $\partial_{\bm{k}}f_{a,\bm{k}}$, so the integrations in $\gamma^{\left(n\right)}_{il}\left(\omega\right)$ with $n=2,3,4$ are over the occupied states. In an insulator under a driving field of finite frequency, $\gamma^{\left(n\right)}_{il}\left(\omega\right)$ with $n=2,3,4$ denote the inter-band quantum geometric contributions to the SHH conductivity and are generally nonzero. Importantly, near the inter-band resonance $2\hbar\omega=\Delta$ ($\Delta$ denotes the band gap), $\gamma_{il}\left(\omega\right)$ in an insulator shows a substantial increase, since the (virtual) inter-band transitions get predominantly enhanced and contribute significantly to the SHH response. In an insulator, such nonvanishing SHH response near inter-band resonance actually manifests the breakdown of Kleinman's conjecture~\cite{Kleinman, Boyd, Cotter}.

\begin{figure*}[t]
\centering
\includegraphics[width=0.96\textwidth]{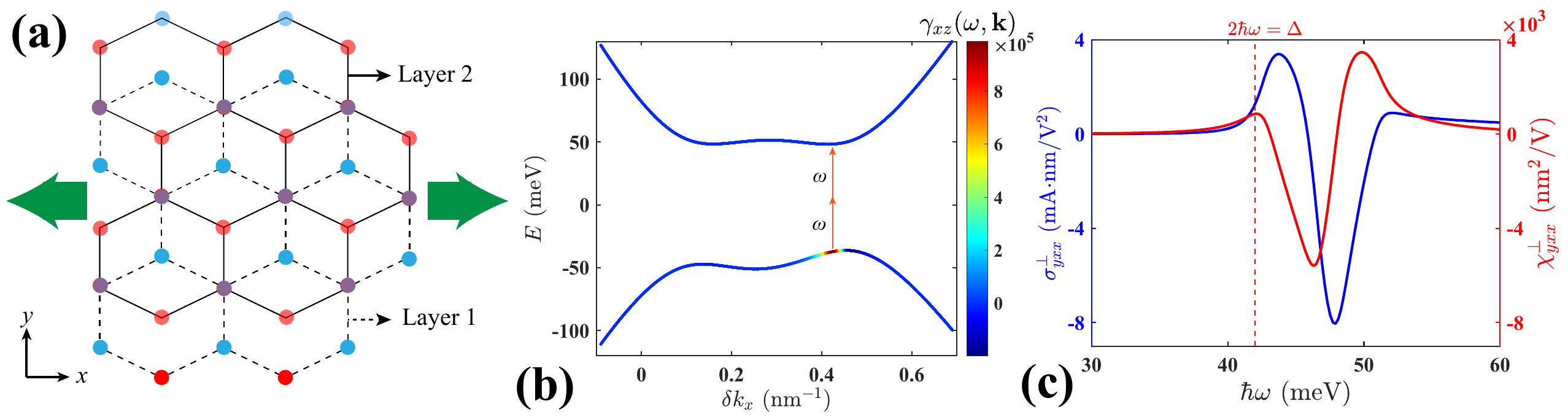}
\caption{The SHH response in an insulating strained Bernal bilayer graphene of C$_{1v}$ symmetry. (a) The lattice structure of Bernal bilayer graphene with uniaxial strain applied along the zigzag direction ($x$-direction). (b) The band dispersions near the $K$ valley ($\delta k_x=k_x-K_x$). A gap of $\Delta=84$ meV is introduced by a $z$-directional electric field. The color variations denote $\gamma_{xz}\left(\omega,\bm{k}\right)$, with $\hbar\omega=42$ meV being the resonant frequency. Here $\gamma_{xz}\left(\omega,\bm{k}\right)$ is in unit of mA$\cdot$nm$^3$/V$^2$. The strain amplitude is $\epsilon=0.03$. (c) The SHH conductivity $\sigma_{yxx}^\perp\left(2\omega;\omega,\omega\right)$ and susceptibility $\chi_{yxx}^\perp\left(2\omega;\omega,\omega\right)$ as functions of the driving frequency. The susceptibility $\chi_{yxx}^\perp\left(2\omega;\omega,\omega\right)$ reaches a local maxima at the resonant frequency, while the local maxima of $\sigma^\perp_{xxy}\left(2\omega;\omega,\omega\right)$ is achieved at a higher frequency. The relative shift of the peaks between $\sigma^\perp_{yxx}\left(2\omega;\omega,\omega\right)$ and $\chi^\perp_{yxx}\left(2\omega;\omega,\omega\right)$ stems from the frequency dependence in the definition: $\sigma^\perp_{yxx}\left(2\omega;\omega,\omega\right)=-i2\omega\epsilon_0\chi^\perp_{yxx}\left(2\omega;\omega,\omega\right)$. The finite $\sigma^\perp_{yxx}\left(2\omega;\omega,\omega\right)$ in the off-resonance regime $\left(2\hbar\omega<\Delta\right)$ arises from the finite relaxation time we introduced in the calculation.}
\label{fig2}
\end{figure*}

\emph{The SHH response and Kleinman's conjecture}.--- In nonlinear optics, Kleinman's conjecture is known as the overall permutation symmetry of the susceptibility indices in the frequency range far below resonances~\cite{Boyd, Cotter}. For the SHG $P_q\left(2\omega\right)=\chi_{qij}\left(2\omega;\omega,\omega\right)\mathcal{E}_i\left(\omega\right)\mathcal{E}_j\left(\omega\right)$, Kleinman's conjecture states that $\chi_{qij}\left(2\omega;\omega,\omega\right)=\chi_{iqj}\left(2\omega;\omega,\omega\right)=\chi_{jiq}\left(2\omega;\omega,\omega\right)=\chi_{qji}\left(2\omega;\omega,\omega\right)=\chi_{ijq}\left(2\omega;\omega,\omega\right)=\chi_{jqi}\left(2\omega;\omega,\omega\right)$ when $\hbar\omega\ll\Delta$. Since the conductivity and susceptibility of the SHG respect $\sigma_{qij}\left(2\omega;\omega,\omega\right)=-i2\omega\epsilon_0\chi_{qij}\left(2\omega;\omega,\omega\right)$ with $\epsilon_0$ being the vacuum permitivity, Kleinman's conjecture seems to preclude any SHH response in insulators when the driving frequency is far off resonance. However, Kleinman's conjecture is only an approximation valid in the far off resonant regime where the frequency dispersion of the media is negligible~\cite{Boyd, Cotter}. As the driving frequency approaches inter-band resonance, (virtual) inter-band transitions get activated and lead to significant frequency dispersions in the susceptibility. Such frequency dispersive properties in the susceptbility breaks Kleinman's conjecture~\cite{Supp} and gives rise to the SHH response in insulators near inter-band resonance. Crucially, in the slightly off-resonance regime, optical absorptions are strongly suppresssed, so the SHH response becomes purely inductive with low energy dissipation.

Since the frequency dependence of the susceptibility near inter-band resonance is universal in insulators, in principle, any insulator that respects the appropriate crystalline symmetry can exhibit a nonzero SHH response. Among the 20 piezoelectric point groups that allow finite SHG~\cite{Boyd, Cotter, Newnham}, by applying Eq. \ref{sigma_H01} to the symmetry analysis in the Supplemental Material~\cite{Supp}, we find that 16 point groups: C$_n$, C$_{nv}$, D$_{n'}$, D$_{2d}$ and S$_4$, with $n=1,2,3,4,6$ and $n'=2,3,4,6$, can support a nonzero SHH response. The specific forms of the SHH susceptbility tensor corresponding to the 16 point groups are presented in Table S3, which provides a clear guideline for seeking the SHH response in real insulators.

\emph{The SHH response in a biased Bernal bilayer graphene under uniaxial strain}.--- To demonstrate the SHH response we propose in insulators, we calculate the SHH conductivity in a biased Bernal bilayer graphene under uniaxial strain. The Bernal bilayer graphene originally respects the D$_{3d}$ point group symmetry. After applying a $z$-directional electric field, the symmetry is reduced from D$_{3d}$ to C$_{3v}$. The $z$-directional electric field introduces a gate tunable band gap~\cite{Heinz, Neto, Yuanbo, Macdonald01}, breaks the inversion symmetry, and enables the SHG in the biased Bernal bilayer graphene~\cite{Pedersen, Marc}. Importantly, if a uniaxial strain is applied along the zigzag direction of the biased Bernal bilayer graphene (Fig. \ref{fig2} (a)), its C$_{3v}$ symmetry is further reduced to the C$_{1v}$ that allows a finite SHH response. Recently, the nonlinear Hall effect has been observed in the metallic regime of a biased bilayer graphene under strain~\cite{TseMing, Ortix02}. It will be interesting to check the nonlinear Hall response when the chemical potential is gate tuned to lie inside the gap.

Our simulation results are shown in Fig. \ref{fig2} (b) and (c), where the bias is set to induce a band gap $\Delta=84$ meV, the chemical potential lies inside the gap, and the uniaxial strain $\epsilon=0.03$ is applied along the zigzag direction. Fig. \ref{fig2} (b) shows the band dispersions near $K$, where the color variations in the bands denote the $\bm{k}$-space distribution of $\gamma_{xz}\left(\omega,\bm{k}\right)$. Here $\gamma_{xz}\left(\omega,\bm{k}\right)$ is the integrand of $\gamma_{xz}\left(\omega\right)$. The nonzero values of $\gamma_{xz}\left(\omega,\bm{k}\right)$ are mainly concentrated around the gap opening region (Fig. \ref{fig2} (b)), indicating that the resulting SHH response is gate tunable. In Fig. \ref{fig2} (c), it is clear to see that both $\sigma^\perp_{yxx}\left(2\omega;\omega,\omega\right)$ and $\chi^\perp_{yxx}\left(2\omega;\omega,\omega\right)$ vanish when $\hbar\omega\ll\Delta$. In the frequency range slightly below resonance, the susceptibility $\chi^\perp_{yxx}\left(2\omega;\omega,\omega\right)$ becomes significantly nonzero, and it reaches a local maxima at the resonance $2\hbar\omega=\Delta$. The significantly nonzero $\chi^\perp_{yxx}\left(2\omega;\omega,\omega\right)$ with $\hbar\omega\in\left[36,42\right)$ meV is purely inductive and involves no energy dissipation. In the frequency range $2\hbar\omega>\Delta$, inter-band transitions occur and give rise to the nonzero $\sigma^{\perp}_{yxx}\left(2\omega;\omega,\omega\right)$. It is worth noting that the Mexican hat shape of the band dispersions in Fig. \ref{fig2} (b) provides 3 characteristic resonant frequencies (see Fig. S3), so $\sigma^\perp_{yxx}\left(2\omega;\omega,\omega\right)$ and $\chi^\perp_{yxx}\left(2\omega;\omega,\omega\right)$ in Fig. \ref{fig2} (c) show 3 local optimal values. More analysis about the frequency dispersions of $\sigma^\perp_{xxy}\left(2\omega;\omega,\omega\right)$ and $\chi^\perp_{xxy}\left(2\omega;\omega,\omega\right)$ can be found in the Supplemental Material~\cite{Supp}. Notably, the local maxima of $\chi^\perp_{yxx}\left(2\omega;\omega,\omega\right)$ at $\hbar\omega=42$ meV reaches 800 nm$^2$/V, which is 2 orders larger than the SHG observed in the two dimensional transition metal dichaocogenides~\cite{Xiaodong, Soavi}. This indicates that the SHH response in the bias gapped Bernal bilayer graphene under uniaxial strain is, in principle, measurable in the Terahertz range.

\begin{figure}[t]
\centering
\includegraphics[width=0.5\textwidth]{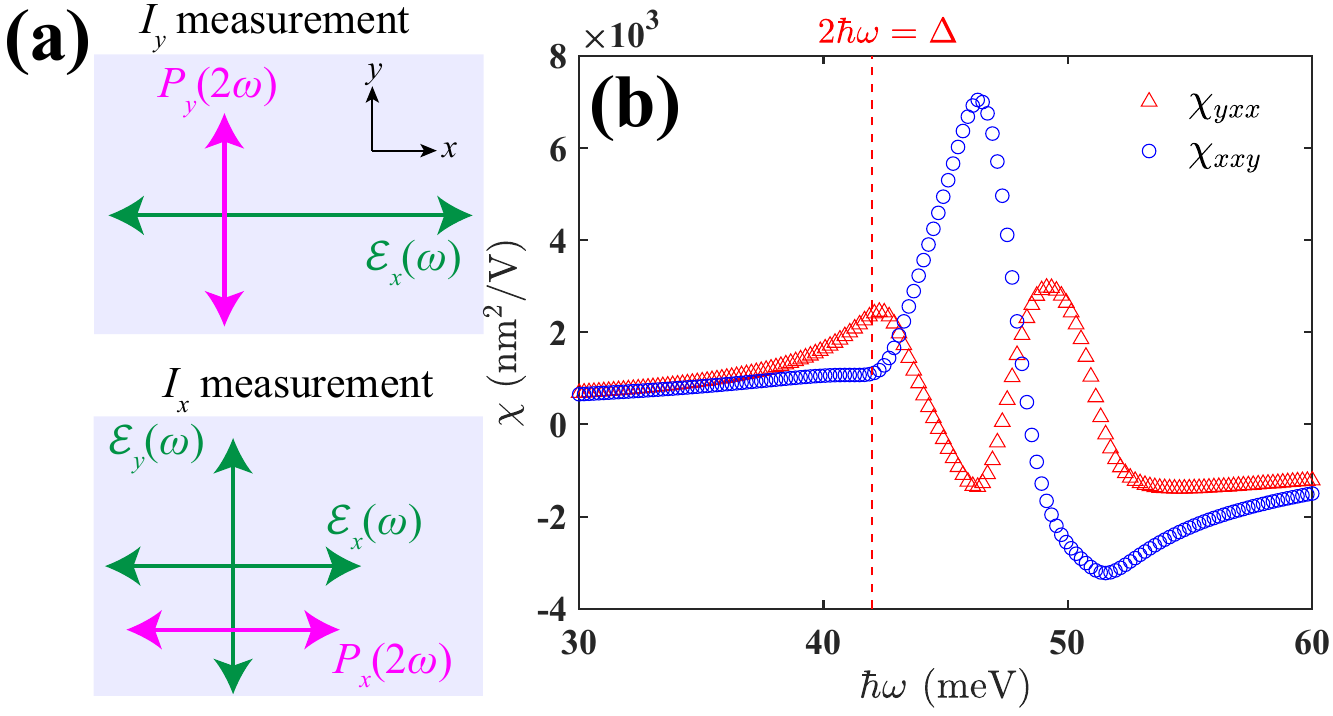}
\caption{Detection of the SHH response. (a) The schematic plot of the SHI measurement. The upper and lower panels correspond to the SHI measured in the $y$ and $x$ directions respectively. Here $I_x$ and $I_y$ denotes the SHI measured in the $x$ and $y$ directions, respectively. (b) The SHH susceptibility as a function of $\omega$. In the far off-resonant regime, $\chi_{yxx}$ and $\chi_{xyy}$ are approximately the same value. As the driving frequency approaches inter-band resonance ($2\hbar\omega\rightarrow\Delta$), $\chi_{yxx}$ and $\chi_{xxy}$ start to differ significantly, which manifests the SHH response.} 
\label{fig3}
\end{figure}

To verify the SHH response in a material, one crucial step is to identify the anti-symmetric property of the nonlinear electric susceptibility (or conductivity). For the biased Bernal bilayer graphene under uniaxial strain in Fig. \ref{fig2} (a), its C$_{1v}$ symmetry dictates that the SHG takes the form~\cite{Supp}
\begin{align}
P_x\left(2\omega\right)=&\epsilon_0\chi_{xxy}\left(2\omega;\omega,\omega\right)2\mathcal{E}_{x}\left(\omega\right)\mathcal{E}_{y}\left(\omega\right),\\\nonumber
P_y\left(2\omega\right)=&\epsilon_0\chi_{yxx}\left(2\omega;\omega,\omega\right)\mathcal{E}_x^2\left(\omega\right)\\
&+\epsilon_0\chi_{yyy}\left(2\omega;\omega,\omega\right)\mathcal{E}_y^2\left(\omega\right).
\end{align}
The nonvanishing Hall component of the SHG arises from the fact that $\chi_{xxy}\left(2\omega;\omega,\omega\right)\neq\chi_{yxx}\left(2\omega;\omega,\omega\right)$ as $2\hbar\omega\rightarrow\Delta$. In practice, one can first identify the zigzag and armchair directions through the polarization resolved second harmonic microscopy~\cite{HuiZhao, Paula, YLi}. After that, one can measure the second harmonic intensity (SHI) in the two orthogonal directions~\cite{Soavi}: 1) $P_y\left(2\omega\right)=\epsilon_0\chi_{yxx}\left(2\omega;\omega,\omega\right)\mathcal{E}^2\left(\omega\right)$ induced by $\bm{\mathcal{E}}\left(\omega\right)=\left[1,0\right]\mathcal{E}\left(\omega\right)$, and 2) $P_x\left(2\omega\right)=\epsilon_0\chi_{xxy}\left(2\omega;\omega,\omega\right)\mathcal{E}^2\left(\omega\right)$ induced by $\bm{\mathcal{E}}\left(\omega\right)=\frac{1}{\sqrt{2}}\left[1,1\right]\mathcal{E}\left(\omega\right)$, as schematically shown in Fig. \ref{fig3} (a). The SHIs in the two directions are $I_x\propto P_x^2$, $I_y\propto P_y^2$, so the difference between $I_x$ and $I_y$ provides direct evidence of the SHH response. In Fig. \ref{fig3} (b), the second harmonic susceptibility $\chi_{yxx}$ and $\chi_{xxy}$ are plotted as functions of $\omega$. Consistent with our analysis, as $2\hbar\omega\rightarrow\Delta$, the deviation between $\chi_{yxx}$ and $\chi_{xxy}$ increases substantially, indicating the emergence of the SHH response. It is noteworthy that one can also identify the SHH response in insulators via directly measuring the Terahertz conductivities of the second order~\cite{AGao, FengWang}.

\emph{Discussions}.---In this work, we investigated the SHH response that involves inter-band transitions and pointed out that insulators, as long as their crystalline symmetry allows, can generally exhibit the SHH response at the frequency near inter-band resonance. We found that a series of frequency dependent quantum geometric quantities in the occupied bands play a pivotal role in generating the SHH response in insulators. Crucially, the SHH response in insulators occurs in both regimes of resonance and off-resonance: At resonance, it manifests as a SHH current, while off-resonance, it corrresponds to an inductive SHH polarization. In particular, we performed calculations on the gapped Bernal bilayer graphene with C$_{1v}$ symmetry and demonstrated that the SHH response is pronounced at the driving frequency near resonance. Apart from the gapped Bernal bilayer graphene under uniaxial strain, quite a few two dimensional insulating materials fall into the category where the symmetry allows the SHH response. Candidate materials with remarkable SHH response in the insulating regime could be extended to the Moir\'e heterostructure of bilayer graphene~\cite{Pablo01, Jianming}, atomically thin MoS$_2$ under uniaxial strain~\cite{Jieshan, JieunLee}, bilayer T$_{\textrm{d}}$-WTe$_2$~\cite{Xiaofeng, Fai03}, in-plane ferroelectric monolayer SnTe~\cite{KChang, Hosub}, and ferroelectric NbOI$_2$~\cite{Loh2}.

It is worth noting that the SHH response arising from the inter-band processes in insulators does not rely on the time reversal symmetry. In an insulator with multiple bands, frequency dispersions originate from the mismatch between the driving frequency and the multiple band gaps, so regardless of whether the insulator respects time reveral symmetry or not, Kleinman's conjecture always gets broken near resonance, generating nonvanishing SHH response. Here, we have mainly focused on the time reversal invariant insulators, but generalizing our formalism to the insulators with broken time reversal symmetry~\cite{Binghai1, PanHe02} is straightforward. Importantly, while the linear order intrinsic Hall effect in insulators requires breaking time reversal symmetry~\cite{Landau}, the SHH response emerges as the lowest order transverse response permitted in time reversal invariant insulators.

Finally, we would like to emphasize that insulators are purely inductive when the driving frequency is below resonance. For an insulator, the SHG below resonance corresponds to the forced vibration of the electric insulating ground state. Since the optical absorption is suppressed by the detuning gap, the energy dissipation involved in the SHG is tiny small. As a result, insulators with remarkable SHH response can be utilized to enable efficient generation of transverse second harmonics~\cite{Linden}, which paves the way for new terahertz optoelectronic device of low-dissipation.

\emph{Acknowledgements}.---W.-Y. H. thanks Shiwei Wu and Kian Ping Loh for their insightful and valuable discussions regarding the practical measurement of transverse second harmonic generation . W.-Y.H. acknowledges the support from the National Natural Science Foundation of China (No. 12304200), the BHYJRC Program from the Ministry of Education of China (No. SPST-RC-10), the Shanghai Rising-Star Program (24QA2705400), and the start-up funding from ShanghaiTech University. K.T.L. acknowledges the support of the Ministry of Science and Technology, China, and Hong Kong Research Grant Council through Grants No. 2020YFA0309600, No. RFS2021-6S03, No. C6025-19G, No. C6053-23G, No. AoE/P-701/20, No. 16310520, No. 16307622 and No. 16309223.


\end{document}